\documentclass[10pt,a4paper,twocolumn]{revtex4}
\usepackage{amsmath}
\usepackage{amssymb}
\usepackage{stmaryrd}
\usepackage{marvosym}
\usepackage{graphicx}
\usepackage{url}

\usepackage{color}

\usepackage{endnotes}

\DeclareMathAlphabet{\mathwee}{OT1}{cmss}{m}{sl}

\newcommand{\hide}[1]{\relax}

\newcommand{\Qm}{\ensuremath{Q}}
\newcommand{\fm}{\ensuremath{f}}

\newcommand{\meff}{m_\text{eff}}

\newcommand{\kB}{\ensuremath{k_\mathrm{B}}}

\newcommand{\SiN}{\ensuremath{\mathrm{SiN}}}

\begin{document}

\title{Ultra-coherent nanomechanical resonators\\ via 
soft clamping and dissipation dilution}
\author{Y.~Tsaturyan}
\affiliation{Niels Bohr Institute, University of Copenhagen,  Blegdamsvej 17, 2100 Copenhagen, Denmark}
\author{A.~Barg}
\affiliation{Niels Bohr Institute,  University of Copenhagen, Blegdamsvej 17, 2100 Copenhagen, Denmark}
\author{E.~S.~Polzik}
\affiliation{Niels Bohr Institute,  University of Copenhagen, Blegdamsvej 17, 2100 Copenhagen, Denmark}
\author{A.~Schliesser}
\email[E-mail: ]{albert.schliesser@nbi.dk}
\homepage[Web: ]{http://slab.nbi.dk}
\affiliation{Niels Bohr Institute,  University of Copenhagen, Blegdamsvej 17, 2100 Copenhagen, Denmark}

\begin{abstract}
The small mass and high coherence of nanomechanical resonators render them the ultimate force probe, with applications ranging from biosensing and magnetic resonance force microscopy, to quantum optomechanics. 
A notorious challenge in these experiments is thermomechanical noise related to dissipation through internal or external loss channels.
Here, we introduce a novel approach to defining nanomechanical modes, which simultaneously provides strong spatial confinement, full isolation from the substrate, and dilution of the resonator material's intrinsic dissipation by %
five orders of magnitude.
It is based on a phononic bandgap structure that localises the mode, without imposing the boundary conditions of a rigid clamp. 
The reduced curvature in the highly tensioned silicon nitride resonator enables mechanical $Q>10^{8}$ at $ 1 \,\mathrm{MHz}$, yielding the highest mechanical $Qf$-products ($>10^{14}\,\mathrm{Hz}$)  yet reported at room temperature. 
The corresponding coherence times approach those of optically trapped dielectric particles.
Extrapolation to $4{.}2$ Kelvin predicts $\sim$quanta/ms heating rates, similar to trapped ions.
\end{abstract}

\maketitle

\section{Introduction}

Nanomechanical resonators offer exquisite sensitivity in the measurement of  mass and force.
This has enabled dramatic progress at several frontiers of contemporary physics, such as the detection of individual biomolecules \cite{Arlett2011}, spins \cite{Rugar2004},  and mechanical measurements of quantum vacuum fields \cite{Aspelmeyer2014a}.
Essentially, this capability originates from the combination of two features: first, a low  mass $m$, so that small external perturbations induce relatively large changes in the motional dynamics.
Second, high coherence, quantified by the quality factor $\Qm$, implying that random fluctuations masking the effect of the perturbation are small.
In practice, a heuristic  $\Qm\propto m^{1/3}$ rule, likely linked to surface losses \cite{Ekinci2005}, often forces a compromise, however.

A notable exception to this rule has been reported recently, in the form of highly stressed silicon nitride string  \cite{Verbridge2008} and membrane  \cite{Zwickl2008} resonators, achieving $\Qm\sim10^6$ at MHz-resonance frequencies $\fm$, and nanogram (ng) effective masses $\meff$.
By now it is understood \cite{Unterreithmeier2010a,Schmid2011,Yu2012} that the pre-stress $\bar \sigma$ ``dilutes'' the dissipation intrinsic to the material (or its surfaces), a feat known and applied also in the mirror suspensions of gravitational wave antennae \cite{Gonzalez1994}.
The resulting exceptional coherence has enabled several landmark demonstration of quantum effects with nanomechanical resonators \cite{Thompson2007, Purdy2013, Purdy2013a,Wilson2015,Nielsen2016} already at moderate cryogenic temperatures.

Systematic investigations \cite{Villanueva2014a} of such silicon nitride resonators have identified an upper limit for the product $Q\cdot f< 6 \times 10^{12}\,\mathrm{Hz}\approx \kB T_\mathrm{R}/(2 \pi \hbar)$ for the low-mass fundamental modes, insufficient for quantum experiments at room temperature $T_\mathrm{R}$.
Better $Qf$-products have been reported in high-order modes of large resonators, but come at the price of significantly increased mass and intractably dense mode spectrum \cite{Wilson2009,Chakram2013}.
Consequently, the revived development of so-called trampoline resonators \cite{Kleckner2011} has received much attention recently \cite{Reinhardt2016,Norte2016}.
In these devices, four thin, highly tensioned strings suspend a small, light ($\meff\sim\mathrm{ng}$) central pad.
The fundamental oscillation mode of the pad can (marginally) achieve $Q\cdot f\gtrsim 6 \times 10^{12}\,\mathrm{Hz}$, provided that radiation losses at the strings' clamping points are reduced through a mismatched, i.~e.\ very thick, silicon substrate \cite{Norte2016}.

In this work, we choose a different approach based on phononic engineering \cite{Maldovan2013}.
Our approach not only suppresses radiation to the substrate  \cite{Wilson-Rae2008} strongly, it also enhances dissipation dilution dramatically.
This is because it allows the mode to penetrate, evanescently, into the ``soft'' clamping region, which exhibits a phononic bandgap around the mode frequency  \cite{Patent2016}. 
This strongly reduces the mode's curvature, whose large value close to a rigid clamp usually dominates dissipation if radiation loss is absent \cite{Gonzalez1994,Unterreithmeier2010a,Schmid2011,Yu2012}.

As a result, we obtain $Qf$-products exceeding $10^{14} \,\mathrm{Hz}$ at MHz frequencies, combined with $\mathrm{ng}$-masses---an ideal combination for quantum optomechanics experiments.
Remarkably, to the best of our knowledge, this is the highest room-temperature $Qf$-product of \emph{any} mechanical resonator fabricated to date.
This includes silicon MEMS devices and bulk quartz resonators, which are fundamentally limited to $Qf\lesssim 3\times 10^{13} \,\mathrm{Hz}$ by Akhiezer damping, but also LIGO's mirror suspensions \cite{Braginsky1985,Ballato1994,Lee2009,Cumming2012,Ghaffari2013}.

\section{Key design features}

\begin{figure*}[htb]
\includegraphics[width=.8 \linewidth]{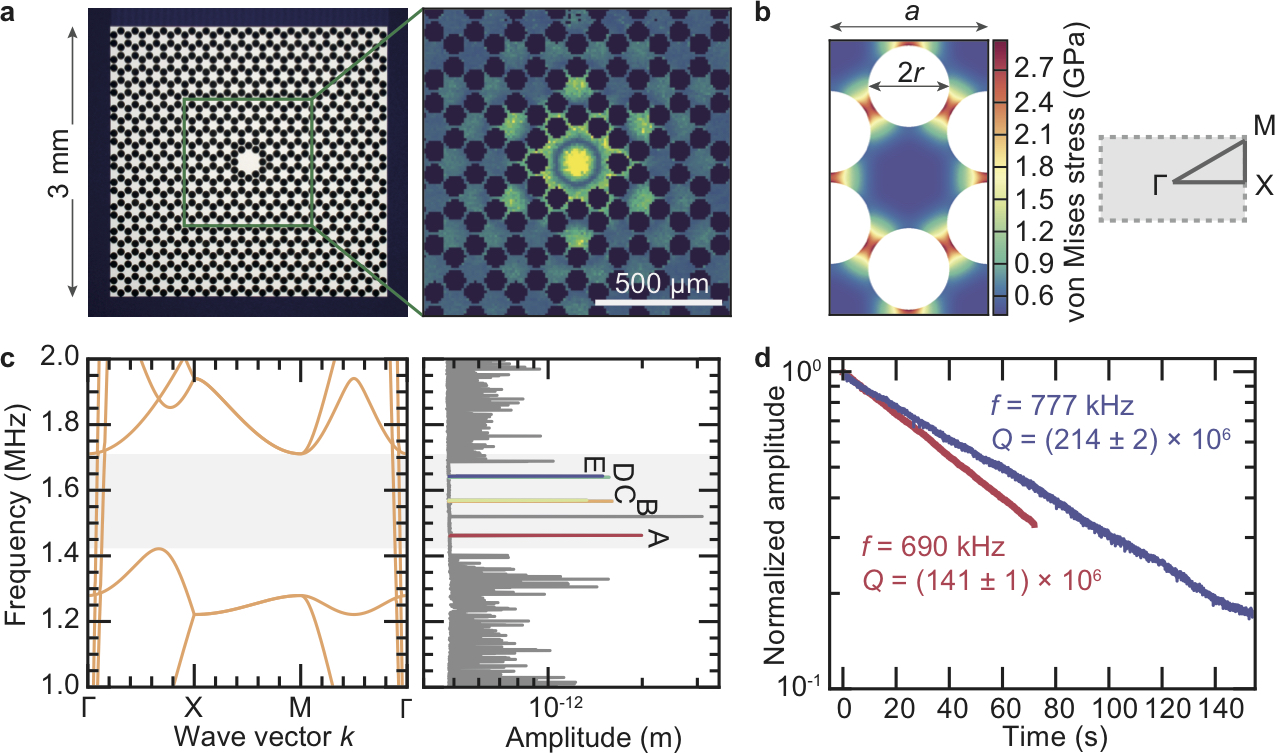}
\caption{
{\bf Device characterisation\, a}) Micrograph of a silicon nitride membrane patterned with a phononic crystal structure (left) and measured out-of-plane displacement pattern of the first localized mode ``$A$'' (right), of a device with lattice constant $a=160\mu \mathrm{m}$. 
{\bf b)} Simulation of the stress redistribution in a unit cell of the hexagonal honeycomb lattice (left) and the corresponding first Brillouin zone (right).
{\bf c)} Simulated band diagram of a unit cell (left) and measured Brownian motion in the central part of the device shown in (a). Localized modes $A$-$E$ are colour-coded,  the peak around $1.5 \,\mathrm{MHz}$ is an injected tone for calibration of the displacement amplitude.
{\bf d)} Ringdown measurements of $A$ (red) and $E$ (blue) modes of two  membrane resonators with $a= 346\mu\mathrm{m}$.
  }
\label{f:features}
\end{figure*}

Figure \ref{f:features} shows the key characteristics of the devices fabricated following this new approach.
A thin (thickness $h=35\ldots240\,\mathrm{nm}$) silicon nitride film is deposited on a standard silicon wafer with a homogenous in-plane stress of $ \sigma\approx1.27\,\mathrm{GPa}$.
The film is subsequently patterned with a honeycomb lattice (lattice constant $a$) of air holes over a $\sim (19 \times 19{.}5) a^2$ square region, where $a=87\ldots346\,\mathrm{\mu m}$ in the batch studied here.
Back-etching the silicon substrate releases membranes of a few mm sidelength (Methods).
Crucially, the  lattice is perturbed by a small number of removed and displaced holes.
They form a defect of characteristic dimension $\sim a$ in the centre of the membrane, to which the mechanical modes of interest are confined.

In contrast to earlier optomechanical devices featuring phononic bandgaps \cite{MayerAlegre2011, Tsaturyan2013,Yu2014, Nielsen2016}, a full bandgap is not expected \cite{Mohammadi2007} here, due to the extreme ratio $h/a\lesssim 10^{-3}$.
A quasi-bandgap can nonetheless be opened \cite{Barasheed2015,Ghadimi2016}, whereby only in-plane modes with high phase velocity persist in the gap (Fig.~\ref{f:features}c).
Under high tensile stress, a honeycomb lattice achieves a relatively large bandgap---about $20\%$ of the centre frequency $ 251\,\mathrm{m/s} \cdot a^{-1}$---with a hole radius $r=0{.}26 a$.
At the same time, the design allows straightforward definition via photolithography, given that the tether width is still above $5\,\mathrm{\mu m}$ even for the smallest $a$.
Evidently, the phonon dispersion is altered dramatically by the in-plane (d.\,c.) stress, which relaxes to an anisotropic and inhomogeneous equilibrium distribution that must be simulated (Fig.~\ref{f:features}b) or measured \cite{Capelle2016} beforehand. 

We characterise the membranes' out-of-plane displacements using a home-built laser interferometer, whose sampling spot can be raster-scanned over the membrane surface (Methods and \cite{Barg2016}).
Figure \ref{f:features}c shows the displacement spectrum obtained when averaging the measurements obtained in a raster scan over a $(200\,\mathrm{\mu m})^2$-square inside the defect, while the ($a=160\,\mathrm{\mu m}$) membrane is only thermally excited.
The spectral region outside  $\sim 1{.}41\ldots 1{.}68\,\mathrm{MHz}$  is characterised by a plethora of unresolved peaks, which can be attributed to modes delocalised over then entire membrane.
In stark contrast, within this spectral region, only a few individual mode peaks are observed, a direct evidence for the existence of a bandgap.
Its spectral location furthermore agrees with simulations to within $2\%$.% (Appendix).

Extracting the amplitude the first peak at $f_\mathrm{A}\approx  235 \,\mathrm{m/s} \cdot a^{-1}$ allows mapping out the (r.\,m.\,s.) displacement pattern of the first mode when raster-scanning the probe (Methods).
Figure \ref{f:features}a (right) shows an image constructed in this way, zooming on the defect region.
The pattern resembles a fundamental mode of the defect, with no azimuthal nodal lines, and its first radial node close to the first ring of holes defining the defect.
Outside the defect, the displacement follows the hexagonal lattice symmetry, but decays quickly with increasing distance to the centre.
This is expected due to the forbidden wave propagation in the phononic bandgap, and leads to a strong localisation of the mode to the defect.

\section{Ultra-high quality factors}

To assess the mechanical quality of the mode, we subject the defect to a second, amplitude-modulated ``excitation'' laser beam, and continuously monitor the defect's motion at the mode frequency, by lock-in detection of the interferometer signal.
When the excitation laser is abruptly turned off, we observe the ring-down of the mechanical mode (Methods).
Under a sufficiently high vacuum ($p\lesssim 10^{-6}\,\mathrm{mbar}$), but at room temperature, the ringdown can last for several minutes at MHz frequencies.
Figure \ref{f:features}d shows an example of an $E$-mode with $f=777\,\mathrm{kHz}$  and amplitude ringdown time $2\tau=(87{.}7\pm\,0{.}8)\mathrm{s}$.
This corresponds to $Q=2\pi f \tau=(214\pm2)\times 10^6$ and $Q\cdot f=(1{.}66\pm0{.}02)\times 10^{14}\,\mathrm{Hz}$.

\begin{figure}[hbt]
\includegraphics[width= \linewidth]{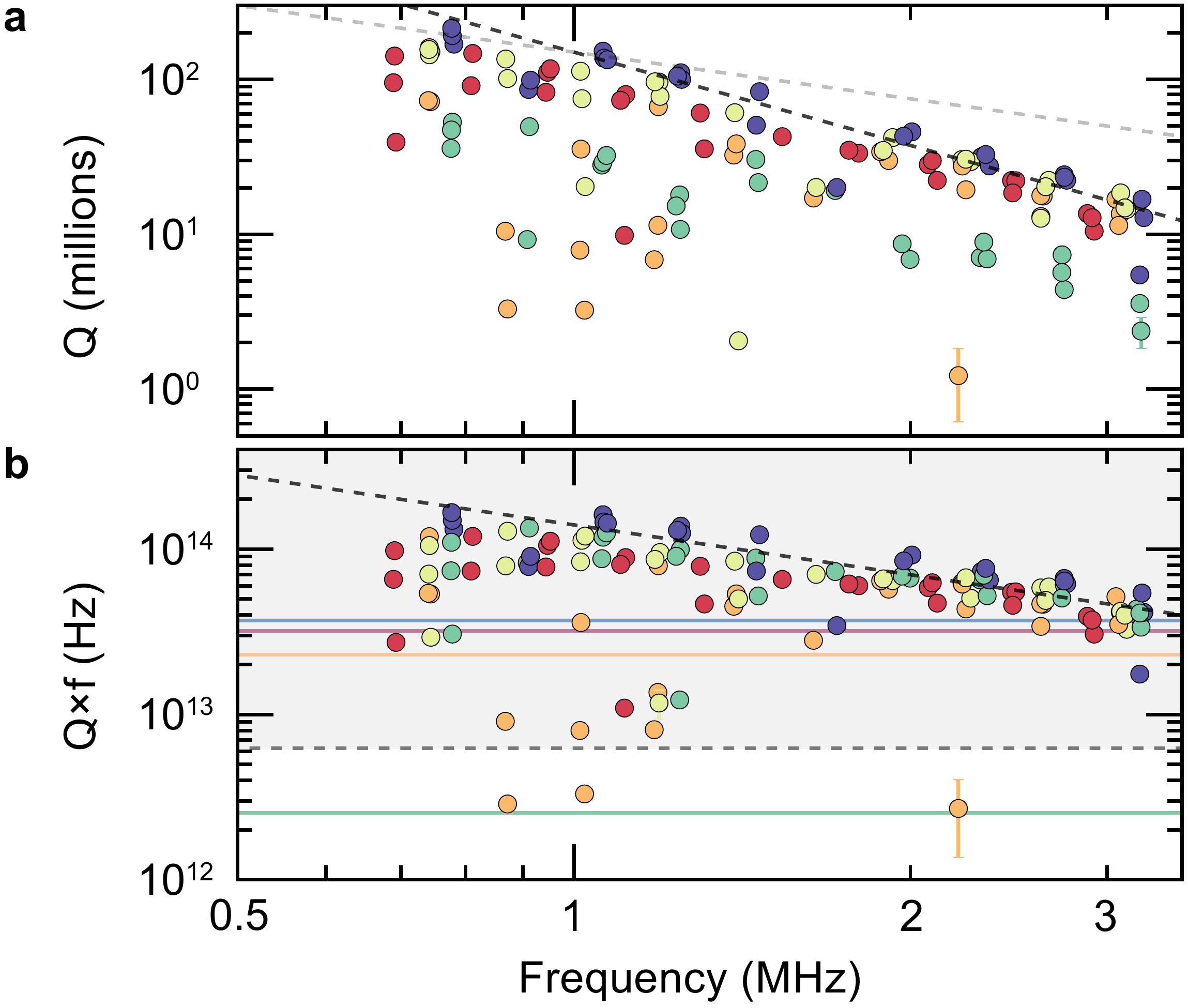}
\caption{
{\bf Quality factor statistics.} 
{\bf a)} Measured $Q$-factors  of $A$-$E$ defect modes in membranes of $h=35\,\mathrm{nm}$ thickness and different size.
Black (grey) dashed line is a $Q\propto f^{-2}$ ($Q\propto f^{-1}$) guide to the eye. 
Colours indicate different localised defect modes, as in Fig. 1c.
{\bf b)} Corresponding $Qf$-products.
For reference, solid  orange, red and blue lines indicate the ``quantum limit''  of crystalline silicon, quartz and diamond resonators, respectively \cite{Ghaffari2013}.
Solid green line shows the expected value for the fundamental mode of a square membrane under $\sigma=1\,\mathrm{GPa}$ stress (\ref{eq:square}), and  
dashed grey line indicates  $Q f= 6 \times 10^{12}\,\mathrm{Hz}$ required for room-temperature quantum optomechanics and reached by trampoline resonators \cite{Norte2016} at $f\approx0{.}2\,\mathrm{MHz}$ (not shown).
}
\label{f:Qf}
\end{figure}

To corroborate and explain this result, we have embarked on a systematic study of more than 400 modes in devices of varying thickness and size (rescaling the entire pattern with $a$).
Figure \ref{f:Qf} shows a subset of quality factors and $Qf$-products measured in 5 different modes of $\sim20$ devices with varying size $a=87\ldots346\,\mathrm{\mu m}$ but fixed thickness $h=35\,\mathrm{nm}$.
Clearly, the $Qf$-products exceed those of trampoline resonators by more than an order of magnitude, reaching deeply into the region of $Q\cdot f> 6 \times 10^{12}\,\mathrm{Hz}$.
It also consistently violates the ``quantum'' (Akhiezer) damping limit of crystalline silicon, quartz and diamond at room temperature, which fundamentally precludes mechanical resonators made from these materials from reaching beyond $Q\cdot f\sim 3\times 10^{13}\,\mathrm{Hz}$ %
\cite{Braginsky1985,Ballato1994,Lee2009,Ghaffari2013}.

Our data, in contrast, do not seem to be limited by Akhiezer damping.
Indeed a crude estimate following  \cite{Ghaffari2013} indicates $Q_\mathrm{Akh}f\sim\mathcal{O}(10^{15}\,\mathrm{Hz})$ for silicon nitride.
Furthermore, since the phonon relaxation times are much faster than the mechanical oscillation period, we would expect constant $Qf$, rather than the $\Qm\propto f^{-2}$ trend discernible in our data.
Thermoelastic damping, another notorious dissipation mechanism in micro- and nanomechanical resonators \cite{Lifshitz2000}, has previously been estimated \cite{Zwickl2008,Chakram2013} to allow $\Qm>10^{11}$ at $f\sim 1\,\mathrm{MHz}$ in highly stressed \SiN{} resonators, and is therefore disregarded.

In absence of radiation loss \cite{Wilson-Rae2008}, stressed membrane resonators are usually limited by internal dissipation.
Its microscopic nature is not known, but evidence is accumulating that it is caused by two-level systems \cite{Faust2013, Yuan2015a} located predominantly in a surface layer \cite{Villanueva2014a}.
Their effect is well captured by a Zener model \cite{Unterreithmeier2010a, Schmid2011, Yu2012}, in which the oscillating strain  ($\tilde \epsilon(t)=\mathrm{Re}[\tilde \epsilon_0 e^{i 2\pi f t}]$) and stress ($\tilde \sigma(t)=\mathrm{Re}[\tilde \sigma_0 e^{i 2\pi f t}]$) fields acquire a phase lag, $\tilde \sigma_0=E \tilde \epsilon_0$, from a relaxation mechanism described by a complex-valued Young's modulus  $E=E_1+i E_2$.
Per oscillation cycle, mechanical work amounting to $\Delta w=\oint \tilde \sigma(t) \dot {\tilde \epsilon}(t) dt=\pi E_2 |\tilde \epsilon_0|^2$ is done in each dissipating volume element. 
Integrating up the contributions yields the loss per cycle $\Delta W=\int \Delta w dV $. 
The comparison with the mode's total energy $W$ determines its quality factor via
\begin{equation}
  Q^{-1}=2\pi \frac{\Delta W}{ W}.
  \label{e:Q}
\end{equation}
In highly stressed strings and membranes, $W$ is dominated by the large pre-stress $\bar \sigma $, counteracting the membrane's elongation.
In contrast, for small amplitudes, the oscillating strain and thus  per-cycle loss is dominated by pure bending.
As a result, $W$ and $\Delta W$ in eq.~(\ref{e:Q}) depend on different parts of the strain tensor $\tilde \epsilon_0=\tilde \epsilon_0^\mathrm{elong}+\tilde \epsilon_0^\mathrm{bend}$ associated with the mode's displacement profile.
For pure out-of-plane displacement $u(x,y)$ of a clamped membrane, this translates into the imperative to minimise bending-related loss
\begin{equation}
  \Delta W\approx\int  \frac{\pi E_2}{1-\nu^2} z^2\left( \frac{\partial^2 u}{\partial x^2}+\frac{\partial^2 u}{\partial y^2} \right)^2 dV
  \label{e:bend}
\end{equation}
over the tensile energy 
\begin{equation}
  W\approx\int \frac{\bar \sigma}{2} \left( \left(\frac{\partial u}{\partial x}\right)^2+\left(\frac{\partial u}{\partial y} \right)^2 \right)dV,
    \label{e:tens}
\end{equation}
where $\nu$ is Poisson's ratio.

\begin{figure*}[bth]
\includegraphics[width= \linewidth]{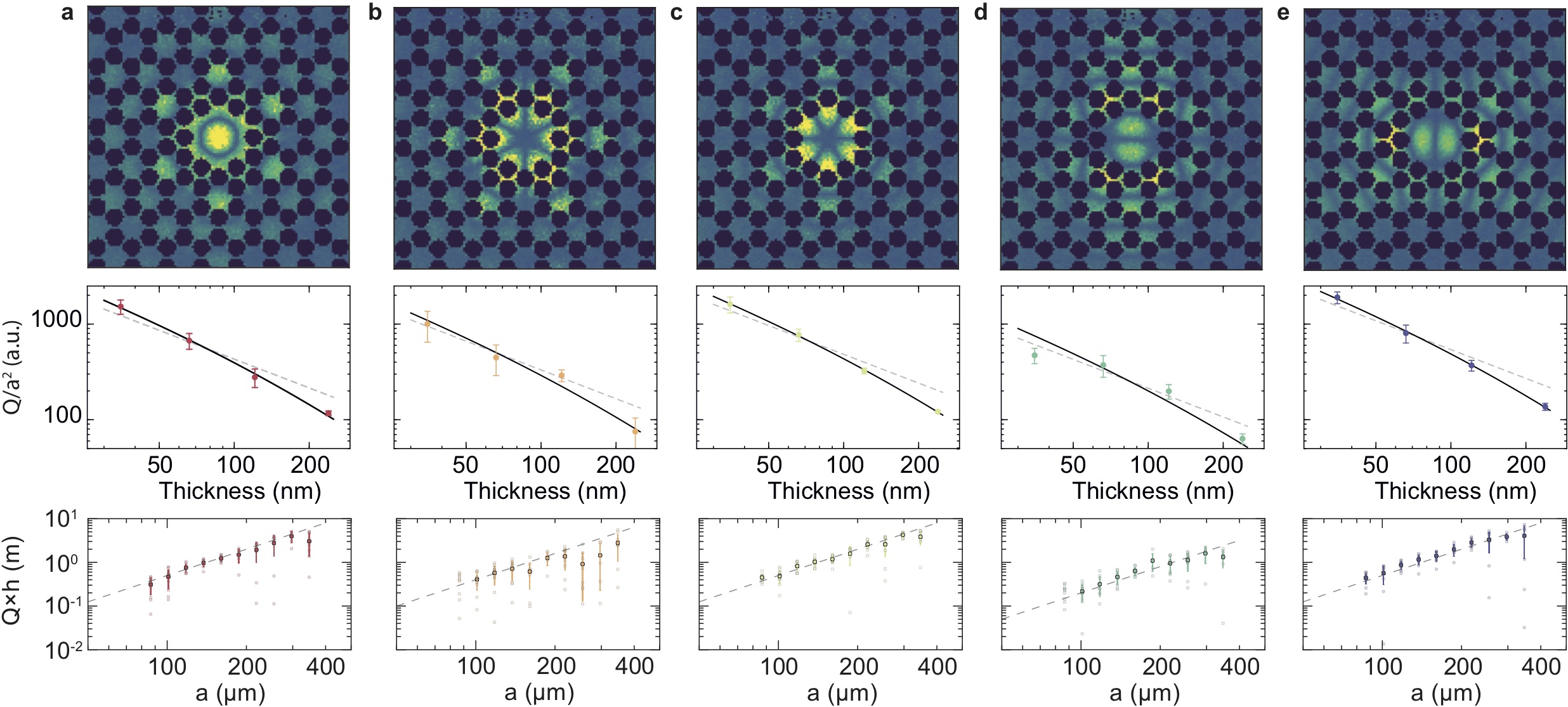}
\caption{{\bf Scaling of quality factors. (a)-(e)} Measured mode shapes of localised defect modes (top) with frequencies \{$f_A$, $f_B$, $f_C$, $f_D$, $f_E\} = \{1.4627,\, 1.5667,\, 1.5697,\, 1.6397,\, 1.6432\}\,\mathrm{MHz}$ for a device with $a=160\mu \mathrm{m}$, as well as characteristic scalings with the membrane thickness $h$ (middle row) and size, parametrized by the lattice constant $a$ (bottom row), when either are varied.
Dashed grey lines in the middle row indicate a $Q/a^2 \propto h^{-1}$ scaling, while the solid black lines take into account the additional losses due to the bulk following eq.\ (\ref{e:Qint}).
Dashed grey line in bottom row indicates $Q\times h \propto a^2$ scaling. 
The semi-transparent points correspond to the individual membranes, while the solid points  (error bars) indicate their mean value (standard deviation).}
\label{f:modes}
\end{figure*}

For the fundamental mode of a plain square membrane of size $L$, this analysis predicts 
\begin{equation}
  Q_{\square}^{-1}=\left(2\lambda+2\pi^2 \lambda^2 \right)Q_\mathrm{int}^{-1} \approx{2 \lambda Q_\mathrm{int}^{-1}},
  \label{eq:square}
\end{equation}
in very good agreement with available data \cite{Yu2012,Villanueva2014a}.
Here, $ \lambda=\sqrt{{E_1}/{(12 \bar \sigma)}} h/L$ quantifies the ``dilution'' of the intrinsic dissipation $Q_\mathrm{int}^{-1}\equiv E_2/E_1$ by the large internal stress $\bar \sigma$.
That is, $\lambda\ll1$, given the extreme aspect ratio $h/L\sim\mathcal{O}(10^{-4})$ and the Young's  modulus $E_1=270\,\mathrm{GPa}$ and prestress $\bar\sigma=1.27\,\mathrm{GPa}$.
In an extension of this model \cite{Villanueva2014a}, extra loss in a $\delta h$-thick surface layer $E_2(z)=E_2^\mathrm{Vol}+E_2^\mathrm{Surf} \, \theta(|z|-(h/2-\delta h))$ can be mapped on a thickness dependent dissipation
\begin{equation}
  Q_\mathrm{int}^{-1}(h) =Q_\mathrm{int,Vol}^{-1}+(\beta h)^{-1}.
  \label{e:Qint}
\end{equation} 
with $\beta=E_1/(6 \,\delta h\,E_2^\mathrm{Surf})$.
If the latter dominates, it yields a total scaling $Q_{\square}^{-1}\propto h^0/L^1 $ with the geometry of the device.
Our devices, however, follow a rather different scaling (Fig.~\ref{f:modes}), even though they are embedded in square membranes.

In this context, it is important to understand the origin of the two terms in eq.~(4): the first, dominating term is associated with bending in the clamping region, while the second arises from the sinusoidal mode shape in the centre of the membrane \cite{Yu2012}.
The former is necessary to match this sinusoidal shape with the boundary conditions $u(\vec r_\mathrm{cl})=( \vec n_\mathrm{cl}\cdot\vec \nabla) u(\vec r_\mathrm{cl})=0 $, where $\vec r_\mathrm{cl}=( x_\mathrm{cl},y_\mathrm{cl})$ are points on, and $\vec n_\mathrm{cl}$ the corresponding normal vectors to, the membrane boundary.
It requires, in particular, that the membrane lies parallel to the substrate directly at the clamp, before it bends upwards supporting the sinusoidal shape in the centre.
The extent, and integrated curvature of this clamping region are determined by its bending rigidity.

The boundary conditions in our case are dramatically different,
\begin{align}
  u_\mathrm{d}(\vec r_\mathrm{cl})-u_\mathrm{pc}(\vec r_\mathrm{cl})&=0\\
  ( \vec n_\mathrm{cl}\cdot\vec \nabla) \left(u_\mathrm{d}(\vec r_\mathrm{cl})-u_\mathrm{pc}(\vec r_\mathrm{cl})\right)&=0,
\end{align}
requiring only the matching of the defect mode $u_\mathrm{d}$ with the mode in the patterned part  $u_\mathrm{pc}$.
If the phononic crystal clamp supports evanescent waves of complex wavenumber $k_\mathrm{pc}$,  it stands to reason that this ``soft'' clamping can be matched to a sinusoidal mode of the defect, characterised by a wavenumber $k_\mathrm{d} \approx \mathrm{Re}(k_\mathrm{pc})\gg |\mathrm{Im}(k_\mathrm{pc})|$ without requiring significant extra bending.
This eliminates the first term in eq.~(\ref{eq:square}), leaving only the dramatically reduced dissipation
\begin{equation}
  Q^{-1}=\eta{\frac{E}{\bar \sigma}}\frac{h^2}{a^2} Q_\mathrm{int}^{-1}(h),
  \label{eq:sofy}
\end{equation}
dominated by the sinusoidal curvature in the defect (and evanescent waves) $\propto k_\mathrm{d}^{2}\propto 1/a^{2}$, whereby the numerical prefactor $\eta$ depends on the exact mode shape. 
In the surface damping (thin-membrane) limit, we have again $Q_\mathrm{int}^{-1}(h)\approx (\beta h)^{-1}$ and obtain the overall scaling $Q\propto a^{2}/h$.
This is indeed the scaling we observe over a wide range of parameters, in all five defect modes, supporting our argumentation (Fig.~\ref{f:modes}).
For the largest thickness, slightly better agreement is found assuming contributions from volume loss (\ref{e:Qint}), in agreement with expectations.

\section{Simulations}

Finite element simulations (Methods) further support the hypothesis of coherence enhancement by soft clamping.
Simulating the entire structure, including defect and the finite, periodically patterned phononic crystal clamp, produces only five modes with substantial out-of-plane displacements within the bandgap.
Their displacement patterns match the measured ones excellently (SI), and the measured and simulated frequencies agree to within $2\%$. 

\begin{figure}[tb]
  \includegraphics[width= \linewidth]{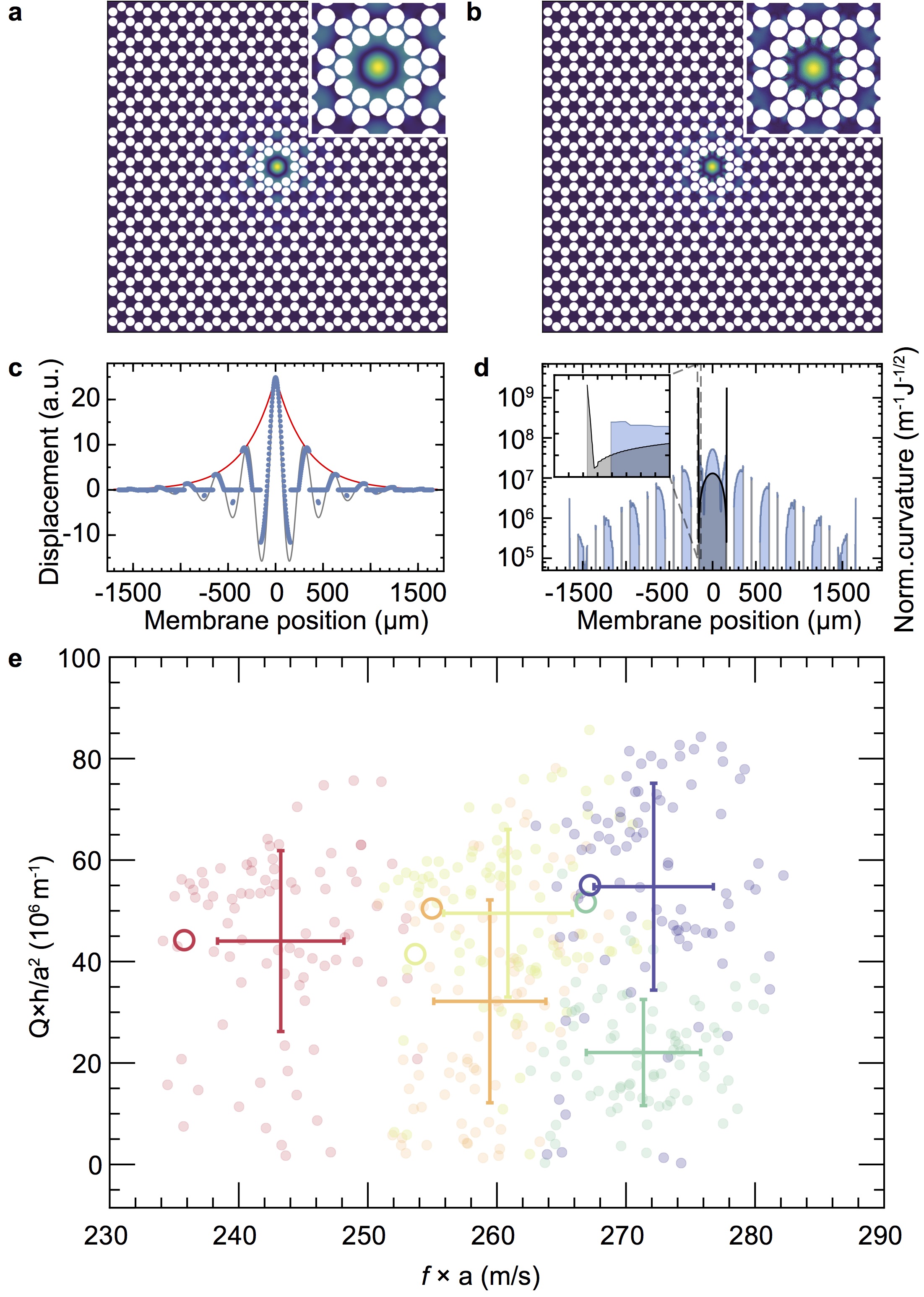}
  \caption{
  {\bf Enhancing dissipation dilution.}
  {\bf a)} Simulated displacement field of the fundamental mode and zoom on the defect (inset). 
  {\bf b)} Absolute value of mode curvature and zoom on the defect (inset). 
  {\bf c)} Simulated displacement along a vertical line through the defect (blue points). 
  The red curve is an exponential function as a guide to the eye, while the grey curve represents a simplistic model of an exponentially decaying sinusoid (see text).
  {\bf d)} Absolute value of mode curvature (blue line) along the same section as c).
  Curvature is normalised to the square-root of the total stored energy in the resonator.
  Also shown, for comparison, is the normalised curvature of a square membrane with the same frequency (grey line).
  Inset is a zoom on the membrane clamp, revealing the exceedingly large curvature of a rigidly clamped membrane, which is absent with soft clamping. 
  {\bf e)} Compilation of measured (transparent markers and errorbars, indicating standard deviations) and simulated (hollow circles) quality factors, normalised to $a^2/h$, consistent with the observed scaling with the corresponding quantities for $h=\left\{35~\mathrm{nm}, 66~\mathrm{nm}, 121~\mathrm{nm}\right\}$.
}
\label{f:simulation}
\end{figure}

Figure \ref{f:simulation} shows the simulated displacement pattern of mode $A$. 
It features strong localisation to the defect and a short, evanescent wave tail in the phononic crystal clamp.

Already a simplistic model $u(0, y)\propto \mathrm{Re}[\exp(i k_\mathrm{pc} |y|)] $, with $k_\mathrm{pc}=2\pi(0{.}57+i\,0{.}085)/a$ reproduces a cross-sectional cut remarkably well (Fig.~\ref{f:simulation}c), supporting the scaling of curvature, and thus damping $\propto 1/a^2$.
Note that more accurate modelling of the Bloch waves, their complex dispersion, and interaction with the defect \cite{Laude2009,Barasheed2015}, is possible, but, in general, has to take bending rigidity into account to obtain the correct mode shape and curvature.

With the full simulated displacements at hand, we are in a position to evaluate the bending energy  (\ref{e:bend}) and the total stored energy (\ref{e:tens}) for a prediction of the quality factor (\ref{e:Q}). 
For computational efficiency, we use the maximum kinetic energy $W_\mathrm{kin}^\mathrm{max}=(2 \pi f)^2 \int \rho u(x,y)^2 dV/2=W$, equivalent to the stored energy (\ref{e:tens}) ($\rho=3200\,\mathrm{kg/m^3}$ is the density of \SiN). 
A comparison of the normalised curvature $|(\partial_x^2+\partial_y^2)u(x,y)|/\sqrt{W}$ 

reveals the advantage of phononic crystal clamping over the fundamental mode of a square membrane: the latter exhibits a 2-order of magnitude larger curvature in the clamping region (Fig.~\ref{f:simulation}d).
The somewhat larger integration domain of the phononic crystal membrane does not overcompensate this, given that the integration is carried out over the mean \emph{squared} curvature (\ref{e:bend}).
Indeed, carrying out the integrals leads to quality factors in very good agreement with our measured values, much higher than the square membranes'.
Figure \ref{f:simulation}e shows the normalised quality factor $Q\times h/a^2$ for the five defect modes of more than $30$ samples, assuming $Q_\mathrm{int}(h=66\,\mathrm{nm})=3750$ \cite{Villanueva2014a}.
Not only are the observed extreme quality factors consistent with simulations, the latter also confirm the trend for the highest $Q$'s to occur in mode $E$, apparent (albeit not very significant) in the measurements.

Not all the modes' measured features are in quantitative agreement with the simulations. 
Small ($<2\%$) deviations in the resonance frequency are likely due to small disagreements between the simulated and fabricated devices' geometry and material parameters, and deemed unproblematic for the purpose of this study.
It is noteworthy, however, that mode $D$ exhibits significantly lower measured quality factors than simulated.
We attribute this to the insufficient suppression of the mode amplitude at the silicon frame, leading to residual radiation losses.
Indeed we observe in simulations that mode $D$ has the largest amplitude at the silicon frame (SI), and the fact that mode $D$ responds most sensitively to the clamping conditions of the sample.

\section{Applications in optomechanics and sensing}

The ultra-high quality factors enabled by soft clamping offer unique advantages for experiments in quantum optomechanics, as well as mass and force sensing.
In quantum optomechanics \cite{Aspelmeyer2014a}, the presence of a thermal reservoir (temperature $T$) has the often undesired effect that it leads to decoherence of a low-entropy mechanical quantum state: for example, a phonon from the environment excites the mechanical device out of the quantum ground state.
This decoherence occurs at a rate
\begin{equation}
  \gamma=\frac{k_\mathrm{B} T}{\hbar Q}=1/\tau
\end{equation}
and sets the timescale $\tau$ over which quantum-coherent evolution of mechanical resonators can be observed.
It is a basic experimental requirement that this time exceeds the oscillation period, so that coherent evolution can be tracked over a number of $\sim 2\pi f \tau>1$ cycles.
At room temperature $T=300\,\mathrm{K}$, this translates to $Q\cdot f>6\times10^{12}\,\mathrm{Hz}$, as already discussed above.
Our measured devices fulfil this condition with a significant margin.

The more challenging requirement  typically is to optically measure and/or prepare  the mechanical quantum state within the time $\tau$.
Since the measurement rate is proportional to the inverse effective mass $1/\meff$, the latter constitutes another important figure of merit.
For a device with $a=160\,\mathrm{\mu m}$, $ h=66\,\mathrm{nm}$, we have measured (Methods) effective masses $\meff$ of
 $\{
4{.}3,\,
4{.}7,\,
4{.}4,\,
9{.}8,\,
7{.}2
\}\cdot(1\pm 0.11)\, \mathrm{ng}$ for the five defect modes, which compare very favourably with $m_\mathrm{eff,\square}= 4{.}9\,\mathrm{ng}$ of 
a square membrane with the same fundamental frequency $f=1.46\,\mathrm{MHz}$ as mode $A$.
Note that we have, in a recent experiment \cite{Nielsen2016}, realised optical measurements on similar square membranes at rates close to  $\Gamma_\mathrm{meas}=2\pi\times 100\,\mathrm{kHz}$, which exceeds $\gamma$ of the new resonators already at room temperature.
In principle, it is thus possible to ground-state cool, or entangle the novel mechanical resonators at room temperature.

The limits in force and mass sensitivity due to thermomechanical noise are also improved by the devices' enhanced coherence and low mass, given the Langevin force noise power spectral density
\begin{equation}
   S_{FF}=2 \meff \frac{2 \pi f}{Q} k_\mathrm{B} T.
\end{equation}

Table \ref{t:parameters} gives an overview of the figures of merit that ensue for the best device we have measured at room temperature.  It also includes an extrapolation of these parameters to liquid helium temperatures.
Here we assumed a \mbox{$2{.}5$-fold} reduction of intrinsic dissipation (\ref{e:Qint}) upon cooling, a factor consistently observed in \SiN{} films \cite{Tsaturyan2013,Faust2013}.
Note that the expected decoherence rates are about an order of magnitude lower than those of optically trapped dielectric particles \cite{Jain2016}, and reach those achieved with trapped ions \cite{Turchette2000}.
It combines with the low effective mass to thermomechanical force noise at the $\mathrm{aN/\sqrt{Hz}}$-level, attractive for force sensing and -microscopy, such as magnetic resonance force microscopy (MRFM) of electron and nuclear spins \cite{Rugar2004,Poggio2010}, as well as mass detection \cite{Hanay2012}.

\begin{table}[htdp]
\caption{Key figures of merit of the $E$-mode in the best ($a=320\,\mathrm{\mu m}$, $h=35\,\mathrm{nm}$) sample at room temperature, where all measurements were performed, and extrapolated to liquid helium temperature.}
\begin{center}

\begin{tabular}{ r l | c | c | l}
  Temperature 			& 	$T	$								& 300			&	$4{.}2$		& K\\
  \hline
  Frequency				&	$f $							& \multicolumn{2}{c|} {777}	& kHz\\
  Effective mass		&	$\meff $							& \multicolumn{2}{c|} {16}& ng\\
  \hline
  Quality factor			&	$ Q $												& 214				&  535		& $10^6$\\
  $fQ$-product			&	$ f\times Q$				& 166				&  416		& THz\\
  Decoherence rate 	& $\gamma/2\pi$ 				& $33{,}000$  	&  $ 175$	& Hz\\
  Coherence time		& $\tau=1/\gamma$		& 5				& 910			&$\mathrm{\mu s}$\\
  \# coherent oscillations
  							& $2\pi f \tau$ 										& 23				& 4400 		& 1\\
  Thermal force noise	& $\sqrt{S_{FF}}$
  																					& 55				& 4.1			& $\mathrm{aN/\sqrt{Hz}}$
  \end{tabular}
  \end{center}
  \label{t:parameters}
\end{table}%

Efficient optical and electronic readout techniques are readily available \cite{Bagci2013, Faust2013}, facilitating also applications beyond cavity optomechanics.
Further, due to the relatively high mode frequencies, $1/f$-type noise, and technical noise, such as laser phase noise, are less relevant.
On a different note, due to the relatively low density of holes, it can be expected that the heat conductivity (provided by unaffected high-frequency phonons) is higher than that of trampoline resonators, an advantage in particular in cryogenic environments, and a fundamental difference to dielectric particles trapped in ultra-high vacuum.
Finally, the sparse spectrum of well-defined defect modes provide an ideal platform for multimode quantum optomechanics \cite{Nielsen2016}, or may  be harnessed for multimode sensing, e.\ g.\ for mass imaging \cite{Hanay2015}.

\section{Outlook}

\begin{figure}[hbt]
\includegraphics[width=.8 \linewidth]{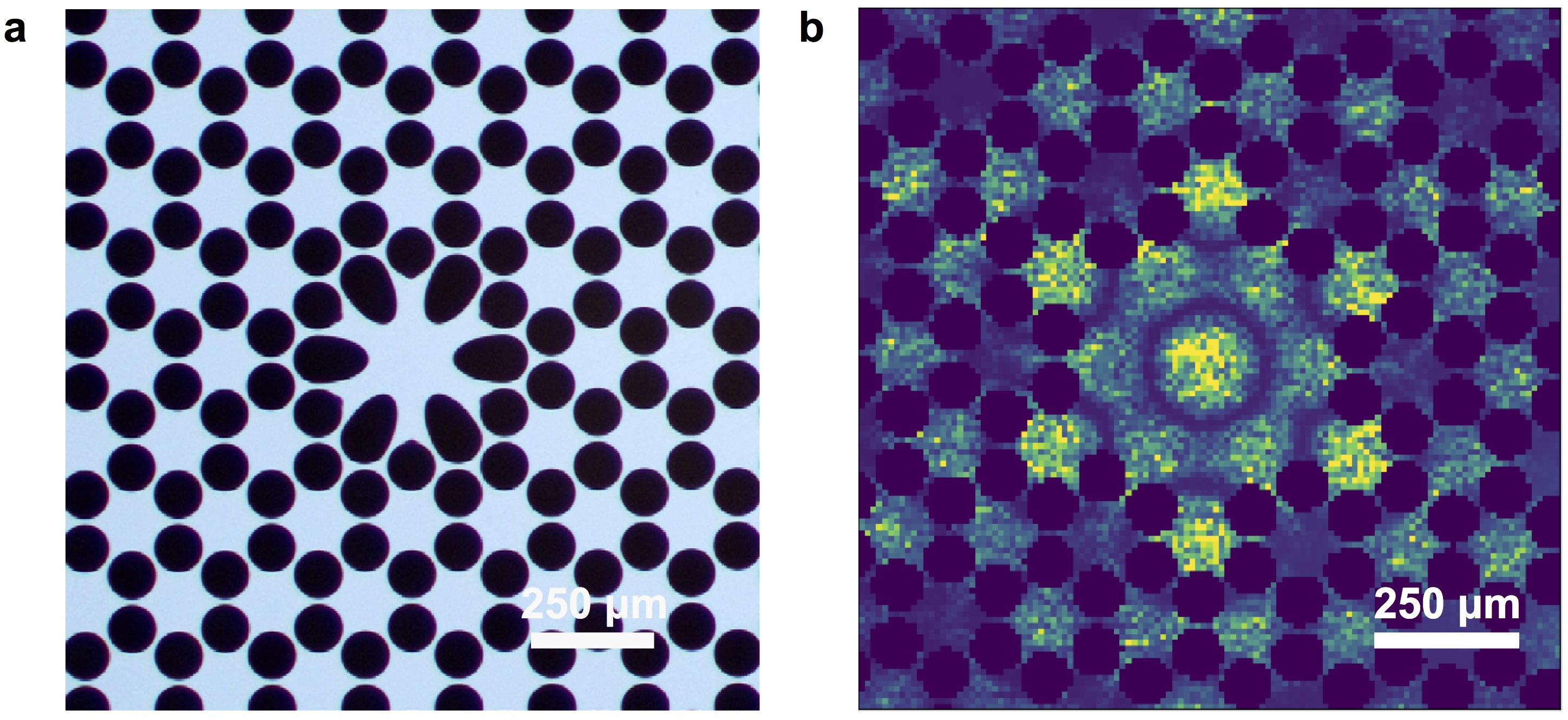}
\caption{{\bf Alternative structures. a)} Micrograph of a trampoline-like resonator embedded in a phononic crystal membrane. {\bf b)} Measured higher-order localised mode of a large defect.}
\label{f:others}
\end{figure}

Clearly, the devices we have discussed above are just specific examples of soft clamping, and many other designs are possible.
Due to the strong suppression of intrinsic dissipation, many other materials can be used to implement high-Q mechanical resonators, including polymers, piezoelectric and crystalline materials, semiconductors, as well as metals.
Engineering of defect shape and size will modify the mode spectrum, mass, and dilution properties, and it is evident that our design can be further optimised, depending on the application. 
For example, larger defects will exhibit a richer multi-mode structure, of interest for multimode optomechanics and mass moment imaging \cite{Nielsen2016,Hanay2015}.
Small, trampoline-like defects have a potential to further reduce mass, as desired for force sensing.
To illustrate this point, Fig.~\ref{f:others} shows two other examples we have realised in our laboratory, and verified to possess a phononic bandgap enhancing dissipation dilution.
Similarly, the phononic crystal clamp can be engineered for stronger confinement, optimised dilution, and/or directional transport.
Exploiting these new opportunities, it will be interesting to apply our soft-clamping approach to truly one-dimensional resonators \cite{Ghadimi2016}, and to explore networks and arrays of defects with ultra-coherent modes with defined couplings.

In summary, we have introduced a novel type of mechanical resonator, which combines soft clamping and dissipation dilution.
Its extremely weak coupling to any thermal reservoir can, on one hand, be harnessed to relax cooling requirements, and thus allow more complex experiments with long-coherence mechanical devices.
On the other hand, if combined with cryogenic cooling, it enables ultra-slow decoherence, which can be overwhelmed even by very weak coherent couplings to other physical degrees of freedom.
A wide range of scientific and technical fields can thus benefit from this new development, including, but not limited to cavity optomechancis \cite{Aspelmeyer2014a,Nielsen2016}, MRFM \cite{Rugar2004,Poggio2010}, mass sensing and imaging \cite{Hanay2012,Hanay2015}, and hybrid quantum systems \cite{Hammerer2009,Kolkowitz2012,Jockel2014,Kurizki2015}, and fundamental studies of quantum mechanics at mesoscopic mass scales.

\section{Methods}
\subsection{Fabrication}

The membrane resonators are fabricated by depositing stoichiometric silicon nitride ($\mathrm{Si}_3\mathrm{N}_4$) via low-pressure chemical vapor deposition (LPCVD) onto a double-side polished $500\,\mu\mathrm{m}$ single-crystal silicon wafer.
 A positive photoresist is spin-coated on both sides of the wafer, and patterns are transferred onto both sides of the wafer via UV illumination, corresponding to the phononic crystal patterns on one side and rectangular patterns on the other side of the wafer. 
 After UV exposure and development of the resist the silicon nitride is etched in these regions using reactive ion etching. The photoresist is removed using acetone and oxygen plasma. In order to protect the phononic patterned side of the wafer during the backside potassium hydroxide (KOH) etch, we use a screw-tightened PEEK wafer holder, only allowing the KOH to attack the side with square patterns.
 Finally, after a 6 hour etch the wafers are cleaned in a piranha solution, thus completing the fabrication process.

\subsection{Characterisation}
Optical measurements of the mechanical motion are performed with a  Michelson interferometer driven by a laser at a wavelength of 1064 nm. We place a sample at the end of one interferometer arm and spatially overlap the reflected light with a strong local oscillator. The relative phase between the two beams is detected by a high-bandwidth ($0-75 \mathrm{MHz}$) InGaAs balanced receiver and recorded with a spectrum analyser. In the local oscillator arm a mirror is mounted on a piezoelectric actuator that follows an electronic feedback from the slow monitoring outputs of the receiver, stabilizing the interferometer at the mid fringe position. Furthermore, the piezo generates a peak with a known voltage and frequency. By measuring the full fringe voltage, the power of this peak is converted into a displacement, which is then used to calibrate the spectrum. Using an incident probe power of $\sim 1 \mathrm{mW}$ the interferometer enables shot noise limited sensitivity of 10 fm/$\sqrt{\text{Hz}}$.

To image mechanical modes the probe beam is focused down to a spot diameter of $2 \mathrm{\mu m}$ and raster-scanned over the sample surface by means of a motorized 3-axis translation stage with a spatial resolution of $1.25 \mathrm{\mu m}$. At each position we extract the amplitude of a few spectral bins around a mechanical peak and thereby construct a 2D displacement map of each mode. The effective masses of mode A-E are extracted from the maximum of the displacement maps after subtracting a background and smoothing. Uncertainties in the mass are based on a $11 \%$ error of the above-mentioned displacement calibration. 

Quality factor measurements are performed by continuously monitoring the membrane motion at a fixed spot on the sample and optically exciting a given mechanical mode using a laser at a wavelength of $880 \mathrm{nm}$ and incident power of $0.5 - 1 \mathrm{mW}$, which is amplitude modulated at the mode frequency using an acousto-optic modulator. We use a lock-in amplifier to analyse the driven motion and record mechanical ringdowns.

For our systematic study of more than 400 mechanical modes, we place a 4-inch wafer each with $\sim$~20 membranes in a high vacuum chamber at a pressure of a few $10^{-7} \mathrm{mbar}$ and gently clamp down the wafer at its rim. We have verified that the mechanical modes with $Q f > 10^{14} \mathrm{Hz}$ are unaffected by viscous (gas) damping to within $10 \%$.

\subsection{Simulations}

We use COMSOL Multiphysics to simulate the phononic crystal patterned membrane resonators. The simulations are typically carried out in two steps. First, we perform a stationary study to calculate the stress redistribution due to perforation, assuming a homogeneous initial in-plance stress $\sigma_{xx} = \sigma_{yy} $. The redistributed stress is subsequently used in an eigenfrequency analysis, where we either calculate the eigenmodes of an infinite array for different wavevectors $\vec k$ in the first Brillouin zone, or simply simulate the eigenmodes of actual devices.

The mechanical quality factors are extracted by calculating the curvature of a given localised mode, which is obtained from an eigenfrequency simulation, as described above. In order to minimize numerical errors, the geometry is densely meshed. We ensure that increasing the number of mesh elements by a factor of ~3 only results in ~10\% change in the integrated curvature.

\bibliographystyle{unsrt}
\bibliography{../pmlit.bib}

\section*{Acknowledgements}

This work was supported by the ERC grants Q-CEOM and INTERFACE, a starting grant from the Danish Council for Independent Research, the EU grant iQUOEMS and the Carlsberg Foundation. 
The authors acknowledge discussions with S.\ Schmid from TU Wien and H.\ Tang from Yale University.
A.~Simonsen provided support with imaging some of the devices.
\\
\section*{Author contributions}

A.~S.\ conceived the idea and directed the research.
A.~S.\ and E.~S.~P.\  provided general research supervision.
Y.~T.\ designed, simulated, and fabricated the samples.
Y.~T.\ and A.~B.\ characterised and imaged the samples.
A.~S.\ and Y.~T.\  analysed the data, developed the model, and wrote the paper.
All authors commented on the manuscript.

\widetext
\clearpage

%%%%%%%%%%%%%%%%%%%%%%%%%%%%%%%%%%%%%%%%%%%%%%%%%%%%%%%%

\setcounter{enumii}{100}%
\setcounter{figure}{0}%
\setcounter{equation}{0}%
\setcounter{section}{0}
\renewcommand \theequation {S\arabic{equation}}%
\renewcommand \thefigure {S\arabic{figure}}
\renewcommand \thesection {S \arabic{section}}

\begin{center}
\large{
\textbf{Supplementary Information: Ultra-coherent nanomechanical resonators\\ via soft clamping and dissipation dilution}}
\end{center}
\vspace{.2in}

\appendix

\section{Simulated displacement}

\begin{figure}[hbt]
\includegraphics[width= \linewidth]{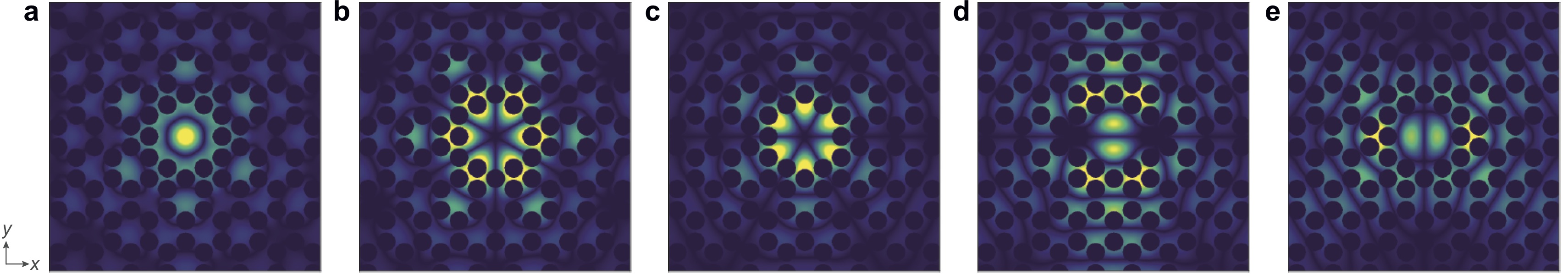}
\caption{{\bf Simulated displacement fields.} Mode shapes of the localised vibrational modes of the defect, showing excellent agreement with the measurements in Fig. \ref{f:modes}.}
\label{f:simulation2d}
\end{figure}

\section{Displacement field projections}

\begin{figure}[hbt]
\includegraphics[width= \linewidth]{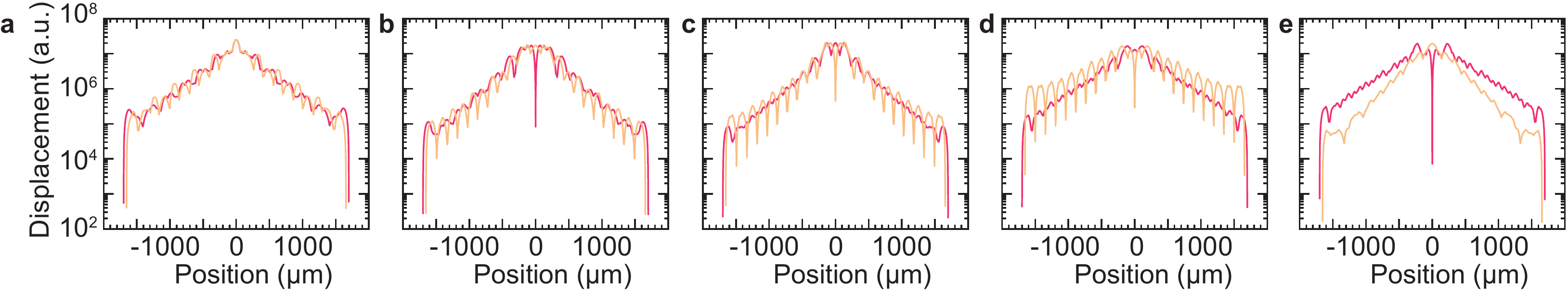}
\caption{{\bf Projections of the displacement fields.} Projections of the simulated displacement fields along the $x$- (orange) and $y$-directions (red) for modes $A-E$.}
\label{f:projections}
\end{figure}

 \end{document}